\documentclass[pra,amsmath,amssymb,twocolumn,floatfix,amsfonts]{revtex4}
\usepackage{graphicx}
\def \beq {\begin{equation}}
\def \eeq {\end{equation}}

\begin{document} 

\title{Towards Quantum Information Processing at the Attosecond Timescale} 

\author{I. K. Kominis$^1$, G. Kolliopoulos$^{1,2}$, D. Charalambidis$^3$ and P. Tzallas$^2$}\email{ptzallas@iesl.forth.gr}
\affiliation{$^{1}$Department of Physics, University of Crete, 71103 Heraklion Greece\\
$^{2}$Foundation for Research and Technology - Hellas, Institute of Electronic Structure \& Laser,
71110 Heraklion Greece\\
$^3$ELI Attosecond Light Pulse Source, ELI-Hu Kft., Dugonics ter 13, 6720 Szeged Hungary}

\begin{abstract}
Coherent processing of quantum information and attosecond science had so far little in common. We here show that recent data in high harmonic emission reveal quantum information processing at the attosecond timescale. By observing the interference pattern created by the spatiotemporal overlap of photons emitted by two interfering electron paths we generate a photon Hadamard gate and thus erase the electron-trajectory information. This allows the measurement of the relative phase in electron-trajectory quantum superpositions and establishes the era of electron-photon quantum coherence and entanglement at the attosecond timescale of high-field physics.

\end{abstract}
\maketitle
Attosecond science is already realizing its promise as a unique quantum microscope of electron quantum dynamics at the atomic unit of time \cite{CK,krausz,dahlstrom,tzallas,goulielm}. In recent years, the spectacular progress in attosecond pulse generation \cite{krausz,agostini,tzallasSpringer} was intimately coupled with the development of new coherent light sources in the EUV spectral region. The physics of high harmonic generation (HHG) has been a central part of these developments. HHG results from the non-linear interaction of intense radiation with matter. Not surprisingly, high-field and ultrafast physics have remained disconnected from another driving force of modern quantum technology, the field of quantum information processing \cite{wineland}. Quantum coherent phenomena have been mostly studied in systems exhibiting long-lived quantum coherence having, until now, little relevance to ultrafast (atto-scale) electron dynamics. Although electron-quantum-path interferences have been studied in HHG \cite{lewenstein,zair}, the vast majority of experimental observables, like HHG spectra and cut-off laws, could be largely understood even within a classical \cite{CK,corkum} or at most a semi-classical model \cite{lewenstein}, without any need to resort to quantum coherence effects or the quantized-radiation formalism \cite{guodrake,eden,eden2,guoPRA1,guoPRA2,varro}, the relevance of which has been debated due to the intense driving laser involved in HHG.

We here argue that new physical observables in HHG \cite{kolliopoulos} can be interpreted in the context of attosecond quantum-information processing. In particular, we interpret the recently observed \cite{kolliopoulos} interference of photons stemming from different electron trajectories with quantum erasure \cite{zeilinger,scully} of electron-trajectory information. Through this quantum erasure we can access the full quantum information imprinted in the electron-trajectory qubit.  We present data connecting the visibility of the photon interference pattern created through this quantum erasure with the properties of coherent superposition of electron trajectories. As a further consistency check of our analysis we unravel the connection of the electron wave packet interference with the properties of the harmonic emission. The theoretical description of electron-photon quantum coherence and entanglement necessitates the quantized-radiation formulation \cite{guodrake,eden,eden2,guoPRA1,guoPRA2,varro} of HHG, along the lines of which we elaborate on the fundamental quantum information on electron-trajectory superpositions that is fundamentally retrievable in HHG.

The process of HHG  \cite{shafir,raz,mairesse} at the single-atom level is governed by the electron quantum path interference \cite{lewenstein,zair}, the properties of which, and hence the properties of the emitted harmonics, strongly depend on the intensity, $I_\ell$, of the driving laser. The intensity $I_\ell$ determines the ponderomotive potential seen by the electrons during their travel in the continuum. In the plateau spectral region there are mainly two \cite{lewenstein} intensity-dependent quantum interfering electron trajectories with different path lengths, the Long (L) and the Short (S) trajectory, that contribute to the off-axis and on-axis harmonic emission \cite{bellini} with phases $\phi^{L}(I_{\ell})$ and  $\phi^{S}(I_{\ell})$, respectively. The simultaneous presence of those trajectories is an impediment to atto-second pulse generation \cite{pulses}, hence so far their spatial distinguishability was taken advantage of in most if not all measurements, and EUV photons were detected from {\it either} the L- {\it or} the S-trajectory. This mode 
of detection, however, automatically eliminates the quantum coherence of-as well as the physically available quantum information imprinted in-the quantum superpositions of S- and L-photons, which naturally arise from the Hamiltonian electron-photon (e-p) interaction during the HHG process. It is the spatio-temporal overlap of S- and L-trajectory photons performed in \cite{kolliopoulos} and the resulting $I_{\ell}$-dependent interference pattern that allows measurement of the phase difference $\Delta\phi^{\rm S,L}=\phi^{S}(I_{\ell})-\phi^{L}(I_{\ell})$. The fundamental reason why this is physically possible is the coherent manipulation of the {\it underlying} quantum coherence of HHG photons as will be explained in the following.
\begin{figure}
\includegraphics[width=8 cm]{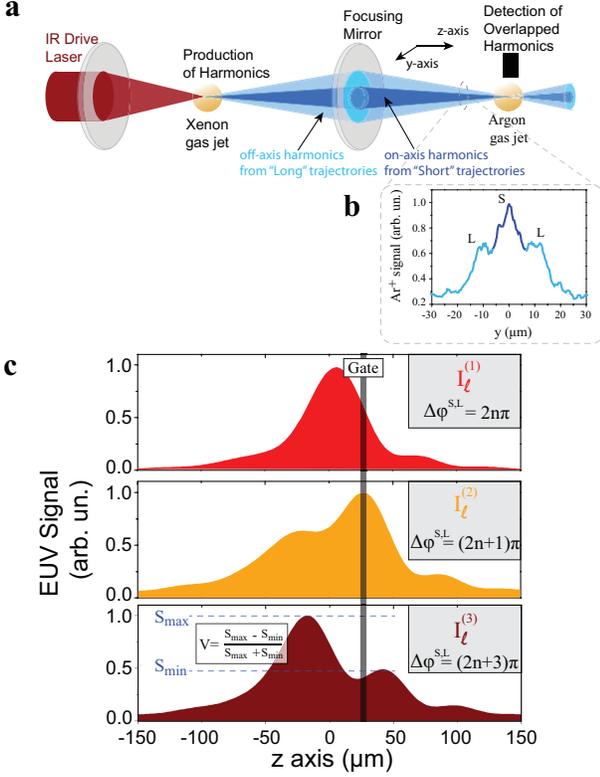}
\caption{(a)  Experimental scheme for realizing a HHG-photon Hadamard gate. The IR drive laser is focused on a xenon gas jet where the HHG photons are produced. On-axis (stemming from S-trajectories) and off-axis (stemming from L-trajectories) photons are then focused onto an argon gas jet, the ions of which are detected. (b) The measured EUV ionization signal along the $y$-axis, depicting the on-axis and off-axis photons. The overlap of the S- and L-peaks is partly due to the projection of the actually well-separated beams on the ion detector. (c) Calculated single-photon ionization signal of the argon jet , proportional to the EUV intensity, as a function of distance along the $z$-axis, shown for three different IR intensities $I_{\ell}$ and the corresponding phase differences $\Delta\phi^{\rm S,L}$. The visibility of the interference pattern is limited by the out-of-plane signal in the projected focus image.}
\label{fig1}
\end{figure}
\begin{figure}
\includegraphics[width=8.5 cm]{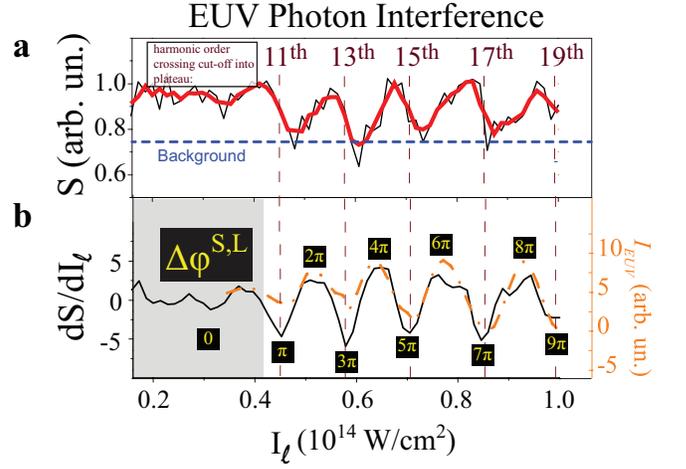}
\caption{(a) Black line: measured EUV signal $S(I_{\ell})$ at $z\approx 20~\mu{\rm m}$ away from focus as a function of $I_{\ell}$. Red line: 3-point average of signal $S(I_{\ell})$ with the thickness being one standard deviation. Blue-dashed line: background stemming from (i) the unfocused EUV beam and (ii) the projected focus image (discussed in  Fig. 1c). Relative to this background, the visibility of the fringes is $V\approx 1$. (b) Derivative $dS/dI_{\ell}$. Maxima (minima) correspond to constructive (distructive) interference of EUV photons with their relative phase $\Delta\phi^{\rm S,L}$ being even (odd) multiples of $\pi$. At low intensities $I_{\ell}$ the harmonics are in the deep cut-off region and the two trajectories degenerate to one with a common phase, hence in the grey-shaded region $\Delta\phi^{\rm S,L}=0$. The orange line (right y-axis) is the calculated EUV intensity at $z=0$. It provides a consistency check in that $dS/dI_{\ell}$ reflects the EUV modulation at $z=0$. The vertical dashed lines are the measured $I_{\ell}$ values at which another harmonic order enters the plateau.}
\label{fig2} 
\end{figure}
\begin{figure*}
\includegraphics[width=15 cm]{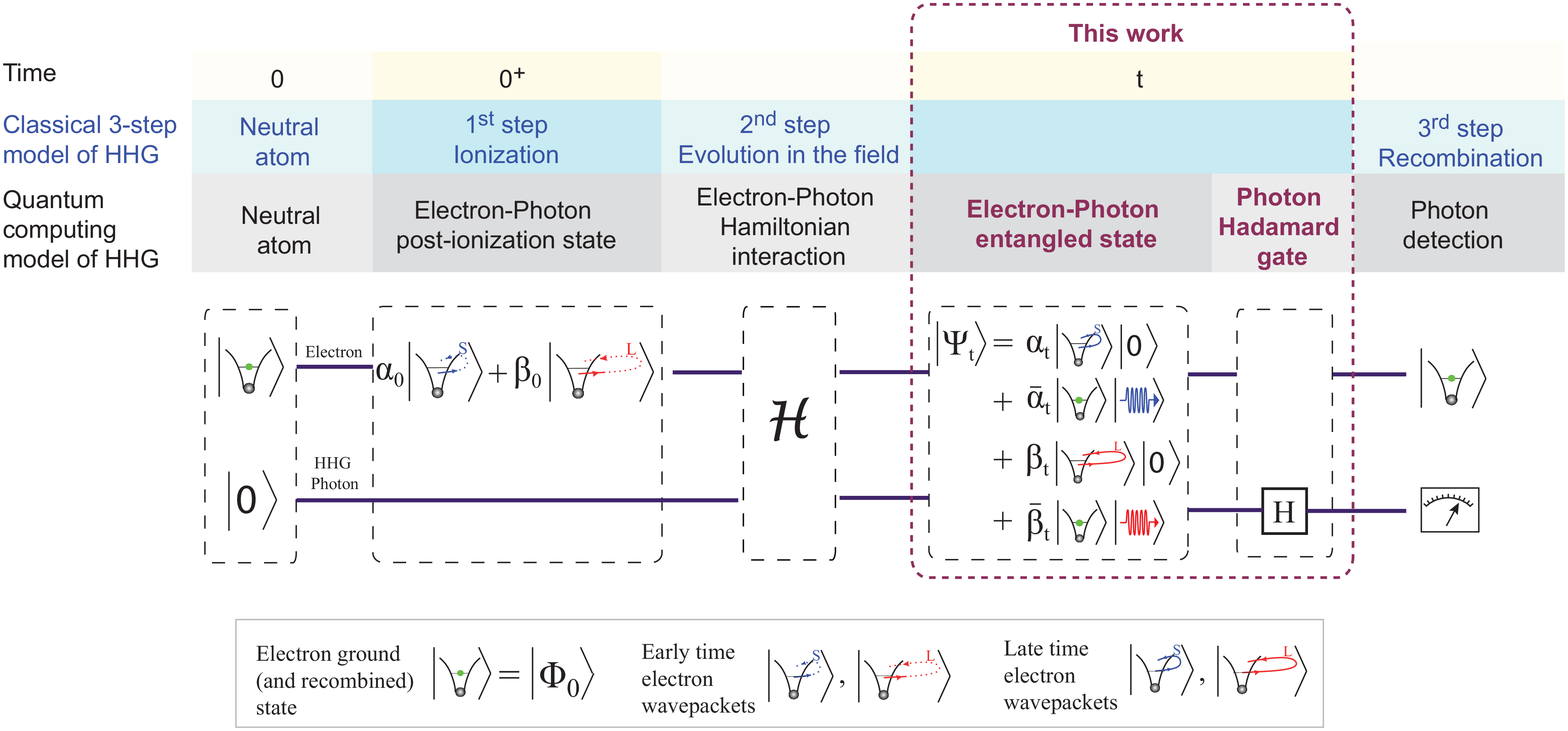}
\caption{The thee-step model of HHG is shown in the context of quantum information processing. The drive IR laser initiates the electronic wave function in a coherent superposition of (L) and (S) trajectories. The initial HHG photon state (for simplicity we omit the drive laser photons) is the vacuum. Due to the Hamiltonian electron-photon (e-p) interaction, the e-p state is entangled at some intermediate time smaller than about 1 fs. The Hadamard gate before the HHG photon detection allows the measurement of the relative phase of the amplitudes $\overline{\alpha}_{t}$ and $\overline{\beta}_{t}$ in the entangled state $|\Psi_{t}\rangle$. The two steps of this work (at time $t$) have no analog in the classical three-step model.}
\label{fig3}
\end{figure*}

In Fig. 1a we depict the experimental scheme used for the data presented here and in \cite{kolliopoulos}. Instead of detecting either S- or L-trajectory HHG photons, both off-axis and on-axis HHG photons  were focused at the ionization detector. In order to visualize the experimental data to be presented in the following and to substantiate the quantum information description of HHG, we calculate the EUV interference pattern along the laser propagation axis (z-axis) for three different drive laser intensities $I_{\ell}$, shown in Fig.1c. The interference maximum oscillates along the z-axis due the $I_{\ell}$-dependence of the phase difference, $\Delta\phi^{\rm S,L}$, between S- and L-trajectory photons. The maximum visibility is obtained at 20 $\mu$m out of focus, and it is there that we place a gate to measure the EUV signal modulation $S(I_{\ell})$, shown in Fig. 2a. However, we can infer the EUV modulation at $z=0$ from the derivative $dS/dI_{\ell}$, since the excursion of the peak in Fig. 1c along the z-axis is 40 $\mu$m. The derivative $dS/dI_{\ell}$ is depicted in Fig. 2b, superimposed with the intensity of the harmonic radiation. The latter is calculated from $I(z=0)=|E_{\rm tot}(z=0)|^2=\sum_{q}|E_{q}^{S}(I_{\ell})+E_{q}^{L}(I_{\ell})e^{-i\Delta\phi^{\rm S,L}(I_{\ell})}|^{2}$, where the sum runs through the detected harmonics $q$ (in our case $q=11,13,15$). To calculate $I(z=0)$ we use the {\it independently measured} values \cite{kolliopoulos} for the L- and S-trajectory electric field amplitudes of each harmonic, $E_{q}^{L}$ and $E_{q}^{S}$, respectively, and the {\it measured} phase difference $\Delta\phi^{\rm S,L}$, linearly interpolated between the minima and maxima of Fig. 2b. We note that the minima (maxima) of the modulation shown Fig. 2b, which are the points of destructive (constructive) interference at the focus $z=0$, are in par with the minima (maxima) of the total harmonic intensity. Importantly, since the electron's ponderomotive energy $U_{p}$ is proportional to the intensity $I_{\ell}$, it is readily found that the beating period of the oscillations in Fig. 2b corresponds to a cut-off energy change by $2\hbar\omega_{\ell}$, i.e. $3.2U_{p}/\hbar\omega_{\ell}=2$, where $\omega_{\ell}$ is the drive-laser frequency. Thus for every two IR photons removed from the drive laser another (odd) harmonic order crosses from the cut-off into the plateau region, demonstrating energy conservation at the quantized electron-photon picture.

The physical basis of our ability to measure $\Delta\phi^{\rm S,L}$ is the access to the entangled e-p state created by the Hamiltonian e-p interaction. In particular, we demonstrate that the classical three-step model of HHG \cite{corkum} is at the fundamental quantum-mechanical level a quantum computation, the diagram of which is shown in Fig. 3. To clearly explain the physics, we consider just one mode for the HHG photons emitted by L- or S-trajectories, call it $|1_{\rm {S}}\rangle$ or $|1_{\rm L}\rangle$, denoting a single photon in either state, while the vacuum is $|0\rangle$.  The initial ($t=0$) and final e-p states in the HHG process are then $|\Psi_{i}\rangle=|\Phi_{0}\rangle\otimes|n_{i}\rangle\otimes|0\rangle$ and $|\Psi_{f}\rangle=|\Phi_{0}\rangle\otimes|n_{f}\rangle\otimes|1\rangle$, respectively, where $\Phi_{0}(\mathbf{r})=\langle\mathbf{r}|\Phi_{0}\rangle$ is the ground-state electron wave function from which the electron is ionized and to which it recombines, $|n_{i}\rangle$ and $|n_{f}\rangle$ are the initial and final photon number states of the IR drive laser, and $|1\rangle$ is the harmonic photon produced, while $|0\rangle$ is the vacuum state. As shown in \cite{eden,eden2,guoPRA1,guoPRA2}, the HHG process can be viewed as a transition from the initial state $|\Psi_{i}\rangle$ to intermediate quantum Volkov states $|\psi_{V}\rangle$, followed by the transition of the latter to the final state $|\Psi_{f}\rangle$, summed over all possible electron momenta $\mathbf{P}$ and photon numbers $n$ of the quantum Volkov state. However, it is known \cite{lewenstein} that among all possible electron paths, there are two that prevail in this sum, namely the S- and the L-trajectory. Hence just after the start of the ionization process (time $t=0^{+}$), the electron and drive laser photon state can be considered to be in a quantum superposition of quantum Volkov states that will evolve into an S- and L-trajectory: 
\beq
|\psi\rangle_{t=0^{+}}=\alpha_{0}|\psi^{\rm S}_{V}\rangle_{t=0}+\beta_{0}|\psi^{\rm L}_{V}\rangle_{t=0},\label{psi0}
\eeq
The amplitudes $\alpha_{0}$ and $\beta_{0}$ depend on experimental parameters such as the laser focus position with respect to the gas jet (and as will be shown next, in our case it is $|\alpha_{0}|\approx|\beta_{0}|$).
The total e-p state will then be 
\beq
|\Psi_{t=0^{+}}\rangle=|\psi\rangle_{t=0^{+}}\otimes|0\rangle,
\eeq
where the HHG photon state is the vacuum.
The e-p Hamiltonian interaction will time-evolve $|\Psi_{t=0^{+}}\rangle$ to the entangled state 
\begin{align}
|\Psi_{t}\rangle&=\alpha_{t}|\psi_{\rm V}^{\rm S}\rangle_{t}\otimes|0\rangle+\beta_{t}|\psi_{\rm V}^{\rm L}\rangle_{t}\otimes|0\rangle\nonumber\\&+\overline{\alpha}_{t}|\Phi_{0}\rangle\otimes|n_{f}\rangle\otimes|1_{\rm S}\rangle
+\overline{\beta}_{t}|\Phi_{0}\rangle\otimes|n_{f}\rangle\otimes|1_{\rm L}\rangle,
\end{align}
which is a coherent superposition of four terms, describing the amplitude for an S- or L-trajectory electron to have (third and fourth terms) or not to have (first and second terms) recombined at time $t$. Clearly, the amplitudes $\overline{\alpha}_{t}$ and $\overline{\beta}_{t}$ (obviously $\overline{\alpha}_{0}=\overline{\beta}_{0}=0$) determine the probability for an S- or L-photon to be 
emitted at time $t$. Their relative phase at time $t$, $\Delta\phi^{\rm S,L}$, reflects, besides the initial relative phase of $\alpha_{0}$ and $\beta_{0}$,  the different phase acquired by the electron along the S- or L-trajectory. Defining $r=|\overline{\beta}_{t}/\overline{\alpha}_{t}|$, the part of $|\Psi_{t}\rangle$ containing one HHG photon will be proportional to 
\beq
|\Psi'_{t}\rangle=|\Phi_{0}\rangle\otimes|n_{f}\rangle\otimes(|1_{\rm S}\rangle+re^{-i\Delta\phi^{\rm S,L}}|1_{\rm L}\rangle)\label{SL}
\eeq
It is clear that if we were to detect {\it either} on-axis {\it or} off-axis photons, i.e. if were we to measure the HHG photon in the $\{|1_{\rm S}\rangle,|1_{\rm L}\rangle\}$ basis, the phase $\Delta\phi^{\rm S,L}$ would not be observable. Now, however, that we apply a Hadamard gate \cite{haroche} before the photon detection, i.e. overlap the on- and off-axis photons and let them interfere, we measure in a different basis spanned by the photon states $|1_{\pm}\rangle=(|1_{\rm S}\rangle\pm|1_{\rm L}\rangle)/\sqrt{2}$, where we assumed for simplicity that our interferometer works like a 50/50 beamsplitter. In terms of $|1_{\pm}\rangle$ we can write 
\begin{align}
|\Psi'_{t}\rangle={1\over\sqrt{2}}|\Phi_{0}\rangle\otimes|n_{f}\rangle
\otimes&\Big[(1+re^{-i\Delta\phi^{\rm S,L}})|1_{+}\rangle\nonumber\\&+(1-re^{-i\Delta\phi^{\rm S,L}})|1_{-}\rangle\Big]
\end{align}
It is now clear that detecting the photon in the $|1_{\pm}\rangle$ basis leads to a $I_{\ell}$-dependent interference pattern of the form $(1+r^{2})+2r\cos\Delta\phi^{\rm S,L}$. Therefore the phase of the interference pattern is the phase difference $\Delta\phi^{\rm S,L}$,  while its visibility $2r/(1+r^2)$ encodes the factor $r$ embodying information about the initial quantum superposition of electron trajectories. The visibility inferred from the interference pattern of Fig. 2a is $V\approx 1$, hence $r\approx 1$. This is consistent with the photon statistics of S-or L-photons. Indeed, from Eq. (\ref{SL}) it follows that if we separately detect either S- or L-photons, their relative rate will be $r^2$. We measured this relative rate and found that $r^2\approx 1.3\pm 0.2$, consistent with the value of $r$ obtained from the visibility of the $I_{\ell}$-interference pattern.
\begin{figure}
\includegraphics[width=7.5 cm]{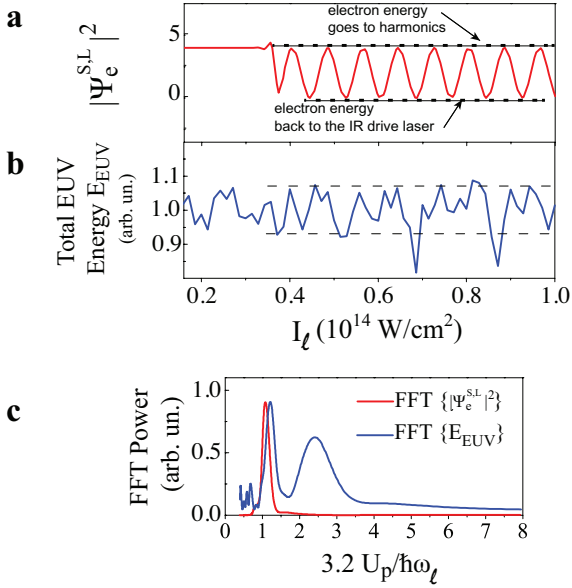}
\caption{ (a) Calculated probability density of two interfering plane waves with $I_{\ell}$-dependent phases. (b) Measured volume-integrated harmonic yield, oscillating in par with the probability density. The pictured modulation sits on top of an $I_{\ell}$-increasing background which has been subtracted away. The maxima (minima) of the beatings (horizontal dashed lines) correspond to the electron energy going to the harmonics (returning to the driving IR field). (c) FFT spectrum (with the x-axis being proportional to the period) of modulations (a) and (b). Both have a component corresponding to one drive-laser photon absorbed, while the modulation of the total harmonic yield has a second component at twice the drive-laser photon energy, due to the generated harmonics entering the plateau. 
} 
\label{fig4} 
\end{figure}

To further elaborate on how the electron-trajectory qubit is imprinted on the HHG photon qubit we plot in Fig. 4a the electron wave packet probability density corresponding to a superposition of the form (\ref{psi0}). Simulating these wave packets with simple plane waves moving just along the polarization of the drive laser (x-axis), the probability density at the nucleus (where recombination takes place) will be $|\Psi_{e}^{\rm S,L}|^{2}=|\alpha e^{-ik_{x}x_{S}(I_{\ell})}+\beta e^{-ik_{x}x_{L}(I_{\ell})}|^{2}$, where $x_{S}(I_{\ell})$ and $x_{L}(I_{\ell})$ are the $I_{\ell}$-dependent path lengths of the S- and L-trajectory, respectively. Since $r\approx 1$, it is $|\Psi_{e}^{\rm S,L}|^{2}\propto 1+\cos(k_{x}\Delta x^{\rm S,L})$, where $\Delta x^{\rm S,L}(I_{\ell})=x_{S}-x_{L}$. It can be shown \cite{kolliopoulos} that $k_{x}\Delta x^{\rm S,L}$ is a linear function of the intensity $I_{\ell}$, and it is through this dependence that the modulation of Fig. 4a is produced. 
The probability is seen to oscillate with a period corresponding to a cut-off energy change by $\hbar\omega_{\ell}$, i.e. $3.2Up_{p}/\hbar\omega_{\ell}=1$, as depicted in the FFT spectrum of Fig. 4c. 
This result further supports the quantized-radiation picture of HHG. Indeed, for yet another drive-laser photon absorbed, an additional momentum $\hbar k_{\ell}$ is transferred to the electrons, and the S-L path length difference $\Delta x^{\rm S,L}$ changes by a de-Broglie wavelength $\lambda_{e}$, leading to a phase difference $\Delta\phi^{\rm S,L}=2\pi$. Thus, the modulation's maxima (minima) correspond to the case when the S- and L-trajectories interfere constructively (destructively) and the electron's kinetic energy goes to the harmonics (back to the drive-laser field). This is corroborated by the {\it measured} harmonic yield (volume-integrated harmonic intensity) shown in Fig. 4b. It's power spectrum (Fig. 4c) also exhibits a peak at $3.2U_{p}/\hbar\omega_{\ell}=1$. Put another way, adding another photon to the drive-laser field, and since even harmonics are ruled our by symmetry, the extra energy either goes to increasing the production rate of odd harmonics or goes back to the field. Incidentally, we also observe a modulation component at  $3.2U_{p}/\hbar\omega_{\ell}=2$, reflecting an additional harmonic order entering the plateau region.This is in agreement with the selection rule for the generation of odd-order harmonics, and consistent with the harmonic cut-off positions discussed in Fig.2. 

Lastly, we have to make two points. (1) Although the produced XUV radiation stems from the coherent contribution of the medium's atoms, our measurement is dominated by the single-atom response. Indeed, the outgoing XUV photon ($q$-th harmonic) wave-vector is $\mathbf{k}_q=q\mathbf{k}_{\ell}+\Delta\mathbf{k}_{g}+\Delta\mathbf{k}_{d}+\nabla\phi_{q}^{\rm S,L}(I_{\ell})$, where $\mathbf{k}_{\ell}$ is the wave-vector of the fundamental. The Gouy phase shift, $\Delta\mathbf{k}_{g}$, and the dispersion-mismatch, $\Delta\mathbf{k}_{d}$, account for the macroscopic propagation effects, while $\nabla\phi_{q}^{\rm S,L}(I_{\ell})$, the phase accumulated by the electrons during their motion in the continuum, represents the single-atom response. The reasons we access the single-atom level are (i) the term $\Delta\mathbf{k}_{d}$ is absent since the harmonic emission time differences are obtained from the measured phase differences between the "S" and "L" trajectories contributing to the {\it same} harmonic generated at the {\it same} EUV frequency, (ii) the term $\Delta\mathbf{k}_{g}$ and the intensity variation of the driving laser field along its propagation in the medium are negligible since the confocal parameter of the IR laser beam is much larger than the medium length, and (iii) we spatially resolve interference maxima and minima resulting from the overlap of the {\it same} frequency of "S"- and "L"-trajectory harmonics. Thus, any intensity variation of the driving field is inconsequential. (2) Although the existence of additional electron trajectories as predicted by TDSE \cite{gaarde} in the case of neon and argon cannot be excluded, the present findings are consistent with the dominance of two electron trajectories. This is because from the linear dependence of $\phi_{q}^{\rm S,L}(I_{\ell})$ on $I_{\ell}$ and the data of Fig. 2 it is found that for $q=11-15$ the proportionality constant $\alpha_{q}^{\rm L}-\alpha_{q}^{\rm S}=40\times 10^{-14}~{\rm rad~cm^2/W}$, which is in agreement with previous experimental findings \cite{heyl} and the theoretical predictions for the dominance of the two shortest electron trajectories \cite{lewenstein,gaarde,varju}. Regarding the connection we make with quantum information processing, the case with more than two trajectories would physically represent not a qubit but a qudit, i.e. a higher-dimensional quantum bit.

In summary, we have experimentally realized basic requirements for attosecond quantum information processing, namely coherent superposition of initial states created by the drive-laser and a single-qubit gate operation, from which we extract the full quantum information of the electron qubit by properly manipulating the photon qubit. This work opens new vistas in the investigation of strong-field light-matter interaction processes leading to coherent EUV light emission \cite{paul,itatani,teubner,ghimire}. Our findings establish the connection
of ultrafast electron-photon dynamics with quantum information processing. 

This work was supported by Laserlab Europe, the Greek
funding program NSRF and by the European Union's Seventh Framework Programme FP7-REGPOT-2012-2013-1 under grant agreement 316165. We also acknowledge B. Bergues, H. Schr\"{o}der and L. Veisz from Max Planck Institut f\"{u}r QuantenOptik for sharing the ion microscope through the DAAD-IKYDA program.

\end{document}